\documentclass[aps,pre,reprint,superscriptaddress,longbibliography,floatfix]{revtex4-2}
\usepackage{amssymb,physics,graphicx,xcolor}
\usepackage{orcidlink}
\usepackage{hyperref}
\hypersetup{
colorlinks=true,
hyperfootnotes=true,
breaklinks=true,
citecolor=blue,
urlcolor=blue,
linkcolor=blue
}

\begin{document}
\title{Optimal quantum transport on a ring via locally monitored chiral quantum walks}

\author{Sara~Finocchiaro\,\orcidlink{0009-0008-3019-6346}}
\affiliation{Center for Nonlinear and Complex Systems, Dipartimento di Scienza e Alta Tecnologia, Universit\`a degli Studi dell'Insubria, Via Valleggio 11, 22100 Como, Italy}
\affiliation{Istituto Nazionale di Fisica Nucleare, Sezione di Milano, 20133 Milano, Italy}

\author{Giovanni~O.~Luilli\,\orcidlink{0009-0008-7624-397X}}
\affiliation{Dipartimento di Fisica, Universit\`a di Milano, 20133 Milano, Italy} 

\author{Giuliano~Benenti\,\orcidlink{0000-0002-0510-0524}}
\email[Contact author: ]{giuliano.benenti@uninsubria.it}
\affiliation{Center for Nonlinear and Complex Systems, Dipartimento di Scienza e Alta Tecnologia, Universit\`a degli Studi dell'Insubria, Via Valleggio 11, 22100 Como, Italy} 
\affiliation{Istituto Nazionale di Fisica Nucleare, Sezione di Milano, 20133 Milano, Italy}

\author{Matteo~G.~A.~Paris\,\orcidlink{0000-0001-7523-7289}}
\email[Contact author: ]{matteo.paris@unimi.it}
\affiliation{Dipartimento di Fisica, Universit\`a di Milano, 20133 Milano, Italy} 
\affiliation{Istituto Nazionale di Fisica Nucleare, Sezione di Milano, 20133 Milano, Italy}

\author{Luca~Razzoli\,\orcidlink{0000-0002-9129-2154}}
\email[Contact author: ]{luca.razzoli@unipv.it}
\altaffiliation[Present address: ]{Dipartimento di Fisica ``Alessandro Volta'', Universit\`a degli Studi di Pavia, Via Bassi 6, 27100 Pavia, Italy; INFN, Sezione di Pavia, Via Bassi 6, 27100 Pavia, Italy.}
\affiliation{Center for Nonlinear and Complex Systems, Dipartimento di Scienza e Alta Tecnologia, Universit\`a degli Studi dell'Insubria, Via Valleggio 11, 22100 Como, Italy} 
\affiliation{Istituto Nazionale di Fisica Nucleare, Sezione di Milano, 20133 Milano, Italy}

\date{\today}

\begin{abstract}
In purely coherent transport on finite networks, destructive interference can significantly suppress transfer probabilities, which can only reach high values through careful fine-tuning of the evolution time or tailored initial-state preparations. We address this issue by investigating excitation transfer on a ring, modeling it as a locally monitored continuous-time chiral quantum walk. Chirality, introduced through time-reversal symmetry breaking, imparts a directional bias to the coherent dynamics and can lift dark states. Local monitoring, implemented via stroboscopic projective measurements at the target site, provides a practical detection protocol without requiring fine-tuning of the evolution time. By analyzing the interplay between chirality and measurement frequency, we identify optimal conditions for maximizing the asymptotic detection probability. The optimization of this transfer protocol relies on the spectral properties of the Perron-Frobenius operator, which capture the asymptotic non-unitary dynamics, and on the analysis of dark states. Our approach offers a general framework for enhancing quantum transport in monitored systems.
\end{abstract}

\maketitle

\section{Introduction}
In a \textit{classical random walk} the walker takes steps in random directions through a network, resulting in diffusive spreading. In a \textit{quantum walk}, instead, the walker takes a quantum superposition of paths, leading to quantum interference effects \cite{venegas2012,portugal2018}. While constructive interference enables ballistic spreading (faster than the classical diffusive spreading), destructive interference can hinder transport or cause localization \cite{keating2007pra}.
Continuous-time quantum walks are versatile for modeling coherent transport in discrete systems \cite{mulken2011}. Equivalent to the tight-binding approximation in solid-state physics and in H\"{u}ckel’s molecular-orbital theory, their applicability is broad across systems
\cite{Li2012scirep,mulken2006jchemphys,Olaya-Castro2008prb,lloyd2011,bose2003prl,rai2008pra,tang2018sciadv}. 
Several studies on environment-assisted quantum transport have highlighted the beneficial interplay of coherent and incoherent dynamics  \cite{mohseni2008environment,plenio2008dephasing,rebentrost2009environment,maier2019prl,Huelga01072013}, where the loss of phase coherence
can enhance transport by suppressing destructive interference \cite{caruso2009highly,Zerah-Harush2018}.
Focusing on purely coherent transport, achieving unit transport efficiency in highly symmetric networks often requires non-trivial delocalized initial states due to symmetry-protected states \cite{varbanov2008pra,razzoli2021entropy}.
Breaking some symmetries, e.g., via network defects \cite{Novo2015scirep,cavazzoni2022pre}, can improve transport.

Chiral quantum walks leverage time-reversal symmetry breaking, typically through complex-valued edge weights \cite{zimboras2013scirep,lu2016pra}, to enable controlled directional transport \cite{frigerio2021,frigerio2022,annoni2024enhanced} or its suppression \cite{Sett2019}, with promising applications in routing \cite{bottarelli2023}. 
Experimentally, chiral quantum walks have been demonstrated using NMR on a 3-qubit system \cite{lu2016pra},
silicon photonic chips \cite{wang2022scichinaphys}, and
gate-based quantum computers \cite{wang2024entropy}, while artificial gauge fields can induce chiral dynamics in ultracold neutral atoms confined in optical lattices  \cite{Aidelsburger2013prl,Tai2017}.
While chirality can mitigate destructive interference by affecting phase coherences, achieving nearly optimal transfer---even on a simple ring \cite{frigerio2022}---often requires fine control of the evolution time \cite{ragazzi25}. This represents an experimental challenge, as it requires a priori knowledge of the optimal time to halt the transfer protocol.
Classically, this relates to first-passage-time problems \cite{redner2001,Bray2013,benichou2014,metzler2014first}, which address the statistics of the time a random walker takes to reach a target. However, in a quantum context the concept of first passage is not meaningful; instead, measurements must be explicitly incorporated \cite{varbanov2008pra}, leading to the notion of first-\textit{detected}-passage time \cite{dhar2015jphysa,dhar2015pra,friedman2017,Thiel2020,kewming2024pra}. 
The latter bridges unitary dynamics and measurement-induced collapse in monitored systems \cite{grunbaum2013}, offering insights into quantum-classical transition and measurement back-action.
Although chirality has been shown to enhance transfer probabilities by lifting energy degeneracies that give rise to \textit{dark states}---initial conditions which, suffering from destructive interference, prevent the detection of the desired state \cite{Thiel2020,wang2024entropy,yin2025}---the optimal interplay between chirality and measurement frequency remains unexplored.

\begin{figure}[!t]
    \centering
    \includegraphics[width=0.8\columnwidth]{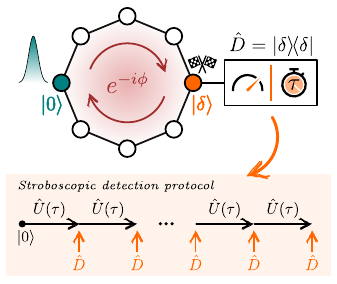}
    \caption{Schematic illustration of the excitation transfer, modeled as a locally monitored chiral quantum walk under a stroboscopic detection protocol, investigated in this work.}
    \label{fig:sketch}
\end{figure}

In this work, we address this gap by investigating a locally monitored chiral quantum walk on a ring (see Fig.~\ref{fig:sketch}), where the purely coherent evolution of the system is repeatedly interrupted by projective measurements at the target (Sec.~\ref{sec:model}). First, we show the effective role of chirality and detection period in enhancing the detection probability at finite time (Sec. \ref{sec:optmz_pbm}). Then, we identify the optimal conditions for maximizing the (asymptotic) detection probability, deriving a general prescription that relies on the Perron-Frobenius spectrum of the non-unitary dynamics (Sec.~\ref{sec:PF_analysis}) and on the analysis of dark states (Sec.~\ref{sec:dark_states}). Finally, we elaborate on the time scale for the asymptotic dynamics to emerge (Sec. \ref{sec:asymptotic}) and conclude with a summary of the main results (Sec. \ref{sec:conclusions}).

\section{Model and first-detected-passage-time problem}
\label{sec:model}
We investigate coherent excitation transfer on a cycle graph with $N$ sites (a ring) under local monitoring (see Fig. \ref{fig:sketch}). The transfer is modeled as a continuous-time chiral quantum walk
and the Hilbert space of the system is conveniently spanned by localized states at the sites of the graph, $\{\ket{j}\}_{j=0,\dots,N-1}$.
In this basis, the Hamiltonian of the chiral quantum walk can be written as \cite{frigerio2021}
\begin{equation}\label{eq:Hamiltonian}
    \hat{H} = \sum_{j=0}^{N-1} e^{-i\phi} |(j+1)_N\rangle\langle j| +  e^{i\phi} |(j-1)_N\rangle\langle j|
\end{equation}
where $(j\pm 1)_N = (j\pm 1) \mod{N}$ due to the periodic boundary conditions.
We set all the diagonal elements to zero to solely focus on the effects of the phase $\phi\in\left[-\frac{\pi}{N},\frac{\pi}{N}\right]$ responsible for the chirality. The upper bound follows from the fact that, in loops, the overall net phase is what matters, and so $-\pi \leq N \phi \leq \pi$ \cite{yang1961prl,frigerio2022}.
Eigenvectors and eigenvalues of the Hamiltonian \eqref{eq:Hamiltonian}, a circulant matrix, respectively read
\begin{equation}
    \ket{\lambda_j}=\frac{1}{\sqrt{N}}\sum_{k=0}^{N-1} e^{i\frac{2\pi}{N}jk}\ket{k},
    \quad 
    \lambda_j=2\cos\bigg(\phi-\frac{2\pi}{N}j\bigg),
    \label{eq:eigvec_eigval}
\end{equation}
with $j=0,\dots,N-1$. Under the time-independent Hamiltonian \eqref{eq:Hamiltonian}, the system initially prepared in a state $\ket{\psi_0}\equiv\ket{\psi(t=0)}$ evolves unitarily according to
\begin{equation}
    \label{eq:free_time_evol}
    \ket{\psi(t)}=\hat{U}(t)\ket{\psi_0}=e^{-i\hat{H}t}\ket{\psi_0},
\end{equation}
where we set $\hbar=1$.

In quantum systems the \textit{first-detected-passage time}, or \textit{hitting time}, is defined through repeated monitoring at the target site \cite{varbanov2008pra,friedman2017}, where the detector reveals the presence or absence of the walker. Here, we assume a stroboscopic detection protocol in which projective measurements are performed at times $\tau, 2 \tau, 3\tau, \ldots$, where the arbitrary finite detection period $\tau >0$ is chosen by the experimentalist. The scheme is illustrated in the box in Fig. \ref{fig:sketch}: starting from an initial state $\ket{\psi_0}=\ket{0}$ localized at a site (without loss of generality we assume the 0th), the monitored dynamics alternates unitary evolution, $\hat{U}(\tau)$ \eqref{eq:free_time_evol}, and projective measurements at the target site, $\hat{D}=\dyad{\delta}$. In the following, we assume the target site to be the opposite to the initial one, i.e., $\ket{\delta}=\ket{N/2}$ for even $N$, and $\ket{\delta}=\ket{(N\pm 1)/2}$ for odd $N$.  
A typical run of the monitored dynamics produces a string of binary measurement outcomes: a sequence of \textit{no} (the walker has not been detected) repeated until a \textit{yes} (the walker has been detected) occurs at time $n \tau$, i.e., at the $n$th detection attempt. This time $n\tau$ is then defined as the first-detected-passage time for the run under investigation.
The probability $F_n$ to first detect the walker at the $n$th attempt is
$F_n=\bra{\theta_n}\hat D\ket{\theta_n}$,
where
$\ket{\theta_n}= \hat{U}(\tau)[(\mathbb{I}-\hat D)\hat{U}(\tau)]^{n-1}\ket{\psi_0}$
is the first-detection (unnormalized) vector \cite{friedman2017}. The detection probability up to time $n\tau$ is
\begin{equation}
    \label{eq:det_prob}
    P_{\rm det}(n)=\sum_{m=1}^n F_m.
\end{equation}
As a final remark, we point out that we investigate local monitoring for excitation transfer 
since purely-coherent transport performs poorly. Indeed, under unitary dynamics, the transfer
probability $P_\delta(t)=|\langle \delta \vert \hat{U}(t) \vert \psi_0 \rangle|^2$ typically 
exhibits low values with rare sharp peaks  
(see Appendix~\ref{app:unitary}).

\section{Optimization problem}
\label{sec:optmz_pbm}
Our purpose is to leverage the key parameters of our model---the detection period $\tau$ between consecutive measurements, and the phase $\phi$ in the Hamiltonian \eqref{eq:Hamiltonian}---to determine an optimal, robust transfer protocol: optimal, as it maximizes the detection probability \eqref{eq:det_prob} over $\phi$ and $\tau$, and robust, as it does not require fine-tuning of parameters. This optimization is subject to a constraint, as we assume the total time $T$ of the process to be a finite resource. Therefore, what we maximize is the detection probability $P_{\rm det}(\phi,\tau)$ after $n$ detection attempts, where $n=\lfloor T/\tau \rfloor$, which is clearly upper bounded by its asymptotic value, $P_{\rm det}(\phi,\tau) \leq P_{\rm det}(n \to \infty)$. In the following, we conveniently set $T=200$, a value which allows us to achieve detection probabilities $> 90\%$ when the number of sites $N$ is relatively small
(see Sec. \ref{sec:asymptotic} for a discussion on the asymptotic time scale),
while keeping the number of measurements $n$ limited. Hereafter, we will focus first on the role of the detection period $\tau$, and then on that of the phase $\phi$.

\begin{figure*}[!t]
    \centering
    \includegraphics[width=0.7\textwidth]{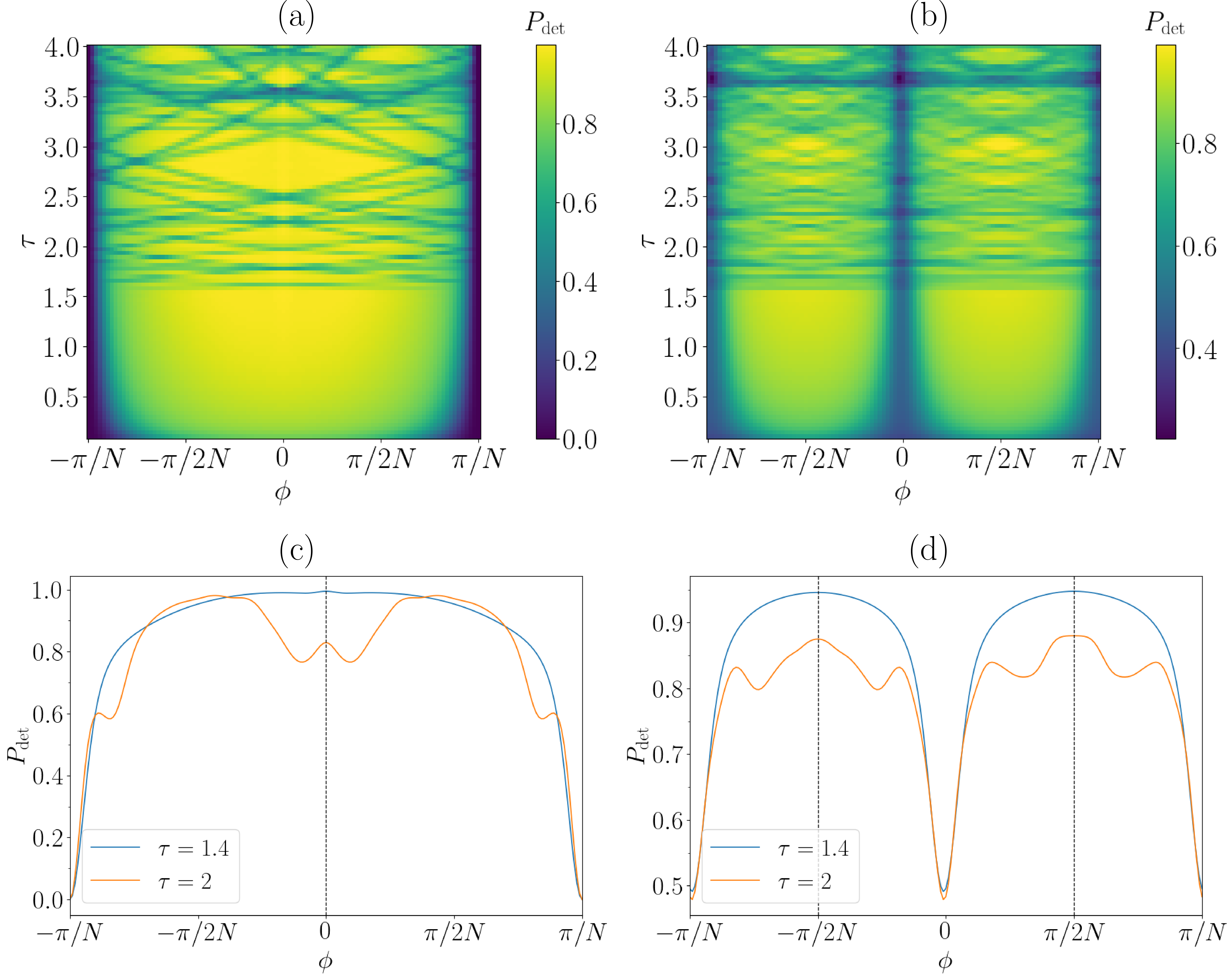}
    \caption{Detection probability as a function of the phase $\phi$ and the detection period $\tau$ at fixed total observation time $T=200$. Density plots $P_{\rm det}(\phi,\tau)$ (a) for $N=20$ and $\delta=N/2$ and (b) for $N=21$ and $\delta=(N-1)/2$. (c,d) Curves $P_{\rm det}(\phi)$ at given $\tau$ for system size $N$ and detection site $\delta$ as in (a,b), respectively.}
    \label{fig:Pdet_density_cuts}
\end{figure*}

\subsection{Optimal detection period $\tau$}
To gain initial insight into how the detection probability depends on the parameters of interest, Fig.~\ref{fig:Pdet_density_cuts} shows the density plots of $P_{\rm det}(\phi,\tau)$ for a fixed number of sites, comparing even ($N=20$) and odd ($N=21$) cases. For both even and odd $N$, the detection probability exhibits two qualitatively distinct regimes separated by a sharp transition at a critical threshold $\tau^*$ in the detection period. Operationally, we define the threshold $\tau^\ast \approx 1.58$ as the first pronounced local minimum in the detection probability that follows the most prominent maximum identified while scanning increasing values of $\tau$.
When $\tau<\tau^*$, $P_{\rm det}(\phi,\tau)$ is smooth, generally increasing with $\tau$ (recall we fixed $T=n\tau$), and remains high over extended intervals. Conversely, when $\tau > \tau^*$, a highly structured pattern with sharp oscillations emerges.
As $\tau \to 0$, we observe a quantum Zeno effect \cite{Facchi_2008,Georgescu2022}. Here, frequent measurements performed on a state $\ket{\delta}$ (orthogonal to the initial one) effectively confine the walker's evolution in the subspace orthogonal to the detection subspace, preventing it from reaching $\ket{\delta}$. As a result, the detection probability approaches zero as $\tau\to 0$. 
Although the detection probability exhibits local maxima for $\tau > \tau^\ast$ that can exceed the maximum observed at $\tau_{\rm opt} \approx 1.53  < \tau^\ast$, 
accessing these higher maxima requires prior knowledge of their locations and fine tuning of $\tau$, conditions that are challenging to meet in practice. Since we aim for a robust and agnostic protocol, we restrict the optimization to the region $0 < \tau < \tau^\ast$, where suitable parameters can always be identified regardless of $N$. In the remainder of this paper, we will show that $\tau^* \to \pi/2^+$ for large $N$, thus providing a well-bounded interval, $0<\tau\leq \pi/2$, in which to maximize the detection probability.

\subsection{Optimal phase $\phi$}
For even $N$ [Fig.~\ref{fig:Pdet_density_cuts}(a)], we observe that for $\tau < \tau^*$ the detection probability is maximized at $\phi_{\rm opt}=0$,
exhibits a lower local maximum at intermediate values $0 < |\phi| < \pi/N$ \footnote{The local maxima at $0 < |\phi| < \pi/N$, which are barely appreciable in Fig. \ref{fig:Pdet_density_cuts}, are discussed in Appendix~\ref{app:PF_even_N}.}, and ultimately vanishes at $\phi = \pm \pi/N$.
Therefore, introducing a nonzero phase in the Hamiltonian (chirality) hinders excitation transfer between opposite sites of the symmetric cycle (even $N$), as it reduces the detection probability at the target. In contrast, chirality enhances this transfer in the asymmetric cycle (odd $N$) [Fig.~\ref{fig:Pdet_density_cuts}(b)]. The directionality induced by the nonzero phases causes the detection probability to increase at either of the two target sites $\delta= (N \pm 1)/2$  opposite to the starting one. In particular, for $\tau < \tau^*$ the detection probability is always maximized at $\phi_{\rm opt}=\pm\pi/2N$, and minimized at $\phi=0, \pm\pi/N$.
Finally, we note that while $P_{\rm det}(\phi, \tau) = P_{\rm det}(-\phi,\tau)$ for even $N$, this symmetry is lifted for odd $N$ [Fig.~\ref{fig:Pdet_density_cuts}(c,d)]. For odd $N$, however, the detection probability is symmetric upon changing both the target site {\em and} the sign of $\phi$.

The numerical results discussed above highlight the presence of optimal parameters, $\phi_{\rm opt}$ and $\tau_{\rm opt}$, 
as well as 
a critical threshold $\tau^\ast$. Below, we present two complementary interpretations of these features based on the analysis of the Perron-Frobenius spectrum and dark states.

\section{Perron-Frobenius analysis}
\label{sec:PF_analysis}
Local monitoring repeatedly interrupts the unitary evolution of the system by projective measurements, resulting in an overall non-unitary evolution. The  elementary step---a free evolution \eqref{eq:free_time_evol} for time $\tau$ followed by the detection attempt $\hat{D}=\dyad{\delta}$---is governed by the (non-unitary) Perron–Frobenius operator \footnote{The authors of Ref.~\cite{Thiel2020} refer to this operator as the survival operator.}
\begin{align}
\label{eq:PF_operator}
\hat{O}(\phi,\tau)
& = [\mathbb{I}-\hat{D}]\hat{U}(\phi,\tau)\nonumber\\
& = \sum_j \mu_j(\phi,\tau) \ketbra{\mu_j(\phi,\tau)}{\bar{\mu}_j(\phi,\tau)}.
\end{align}
Since the operator is non-Hermitian, it admits eigendecomposition in the biorthogonal basis consisting of right and left eigenvectors, $\{\ket{\mu_j},\ket{\bar{\mu}_j}\}$ with eigenvalue $\mu_j \in \mathbb{C}$, such that $\hat{O}(\phi,\tau)\ket{\mu_j}=\mu_j\ket{\mu_j}$ and $\bra{\bar{\mu}_j}\hat{O}(\phi,\tau)=\mu_j\bra{\bar{\mu}_j}$. The orthogonality and completeness relations read $\braket{\bar{\mu}_j}{\mu_k}=\delta_{j,k}$ and $\mathbb{I}=\sum_j \ketbra{\mu_j}{\bar{\mu}_j}$, respectively \cite{Horn_Johnson_2012}. Here and throughout the paper, we omit the explicit dependence of eigenvalues and eigenvectors on $\phi$ and $\tau$ for brevity.
The Perron-Frobenius analysis allows us to characterize the asymptotic dynamics of the system in the long-time limit \cite{Gaspard_1998,Khodas2000,GaciaMata2003,chruściński2025}. Repeatedly applying such operator on an initial state $\ket{\psi_0}$ generates a non-unitary evolution which does not preserve the probability. The \textit{survival probability} 
\begin{equation}
    S(n)
    = \Vert \hat{O}^n(\tau) \vert \psi_0 \rangle \Vert ^2
    = \sum_{j=0}^{N-1}  \vert \mu_{j} \vert^{2n} \langle \psi_0 \vert \mu_j \rangle \langle \bar{\mu}_j \vert \psi_0\rangle
    \label{eq:surv_prob}
\end{equation}
is the probability that the detector has not clicked after $n$ detection attempts---that is, the walker has not been detected yet---and relates to the detection probability \eqref{eq:det_prob} via $S(n)=1-P_{\rm det}(n)$.
The fact that $S(n)\leq 1$ means that $\vert \mu_j \vert \leq 1 \; \forall j$ (see \cite{Thiel2020} for details). In particular, eigenvectors with $\vert \mu_j \vert < 1$ yield exponentially decaying contributions to the survival probability that vanish as $n \to \infty$, while eigenvectors with $\vert \mu_j \vert = 1$ yield non-decaying contributions that remain finite in this limit. The asymptotic dynamics are thus determined by the largest-modulus eigenvalue of the Perron-Frobenius operator $\mu_{\rm PF}$---i.e., the closest to the unit circle---and its corresponding eigenvector(s).
Since by definition $\vert \mu_j \vert \leq \vert \mu_{\rm PF} \vert \leq 1\,\forall j$, if $\vert \mu_{\rm PF} \vert <1$, or if $\vert \mu_{\rm PF} \vert = 1$ and the initial state is orthogonal to the corresponding eigenspace, then the detection probability can approach 1 arbitrarily closely, because the survival probability \eqref{eq:surv_prob} decays exponentially fast with the number of measurement attempts $n$. Otherwise, one has to identify the subleading eigenvalue (i.e., the largest eigenvalue in modulus strictly less than one) whose eigenspace overlaps with the initial state.
In this case, the survival probability attains a finite asymptotic value $S(\infty)>0$, resulting in $P_{\rm det}(\infty)<1$. Minimizing the modulus of the relevant eigenvalue---depending on the scenario---serves two purposes: (i) Identifying parameters, if any, such that $\vert \mu_{\rm PF} \vert < 1$ and (ii) minimizing the number of measurements attempts $n$ required to approach $P_{\rm det}(\infty)$ arbitrarily closely.
The analysis of the eigenproblem of $\hat{O}(\phi,\tau)$ will equip us with the necessary tools to understand both the presence of the threshold $\tau^*$ and the optimal values of the parameters. For clarity of discussion, we focus here on the case of odd $N$, deferring the case of even $N$ to Appendix~\ref{app:PF_even_N}.

\begin{figure}[!t]
    \centering
    \includegraphics[width=\columnwidth]{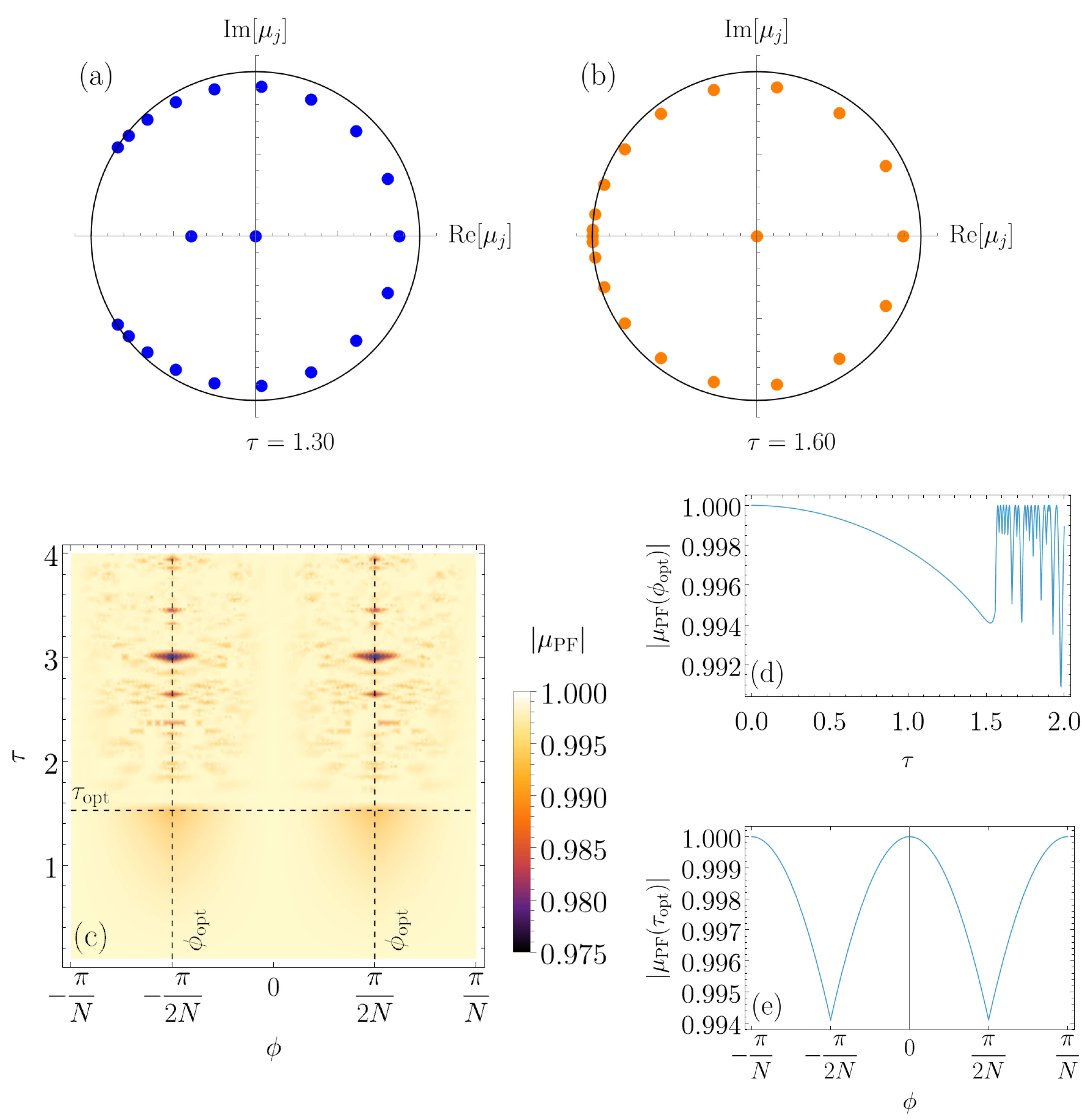}
    \caption{Eigenvalues of the Perron-Frobenius operator \eqref{eq:PF_operator} on the unit circle for two representative values of the detection period, (a) $\tau<\tau^*$ and (b) $\tau>\tau^*$ at $\phi=\pi/2N$, $\delta=10$.
    (c) Density plot of the largest-modulus PF eigenvalue $|\mu_{\rm PF}|$ as a function of $\phi$ and $\tau$. (d) Curve $|\mu_{\rm PF}|$ as a function of $\tau$ at fixed $\phi=\pi/2N$. (e) Curve $|\mu_{\rm PF}|$ as a function of $\phi$ at fixed $\tau=1.53$. Other parameters are: $N=21$, $\delta=10$. 
    }
    \label{fig:maxmodPF_Nodd}
\end{figure}

The existence of the threshold $\tau^*$ is reflected in the Perron-Frobenius spectrum---evaluated at fixed $N$ and $\phi=\phi_{\rm opt}$---which differs between the regimes $\tau < \tau^*$ or $\tau > \tau^*$ [Fig.~\ref{fig:maxmodPF_Nodd}(a,b)]. For $0<\tau<\tau^*$, all the eigenvalues fall within the unit circle, $\vert \mu_j \vert < 1\; \forall j$ [see also the largest-modulus eigenvalue in Fig. \ref{fig:maxmodPF_Nodd}(d)], meaning that all the modes decay and so the excitation will be transferred to the target state with unit probability in the long-time limit, $P_{\rm det}(n \to \infty)=1$.
For $\tau>\tau^*$, instead, there may exist unit-modulus eigenvalues, indicating the emergence of modes which do not decay over time \footnote{Regardless of $\tau$, $\phi$, and $N$ (even or odd), we always observe the presence of a null eigenvalue with corresponding eigenstate $\vert \mu=0\rangle = \hat{U}( \phi,-\tau)\vert \delta \rangle$, which trivially leads to $\hat{O}(\phi,\tau)\vert \mu=0\rangle \equiv 0 $.}.
These results align with the numerical ones discussed above [Fig. \ref{fig:Pdet_density_cuts}(b,d)], including our previous estimate $\tau^*\approx 1.58$.

The optimal parameters, $\tau_{\rm opt}$ and $\phi_{\rm opt}$, can be estimated by minimizing the modulus of $\mu_{\rm PF}$, the eigenvalue governing the asymptotic dynamics in the long-time limit.
The study of $\vert \mu_{\rm PF}\vert$ as a function of $\phi$ and $\tau$ in Fig. \ref{fig:maxmodPF_Nodd}(c) reveals the following. At $\phi = \phi_{\rm opt}=\pm\pi/2N$ (odd $N$), $\vert \mu_{\rm PF} \vert$ decreases smoothly with increasing $\tau<\tau^*$, attains a minimum at $\tau_{\rm opt}$, and for $\tau>\tau^*$ its behavior becomes highly oscillatory, with $|\mu_{\rm PF}|=1$ for specific values of $\tau$ [Fig. \ref{fig:maxmodPF_Nodd}(d)]. At a given detection period $\tau < \tau^*$, $\vert \mu_{\rm PF} \vert$ is minimum at the optimal phase $\phi_{\rm opt}$ [Fig. \ref{fig:maxmodPF_Nodd}(e)].
The locations $(\phi,\tau)$ of the minima (maxima) of $\vert \mu_{\rm PF} \vert$ are consistent with those of the maxima (minima) of the detection probability in Fig. \ref{fig:Pdet_density_cuts}(b,d), supporting the Perron-Frobenius analysis as a valuable tool to optimize the transport.
In the next section, we discuss in terms of dark states the presence of regions where the detection probability is minimum, the emergence of the threshold $\tau^*$ in the detection period, as well as that
$\tau^*\to \pi/2^+$ as $N$ increases.

\section{Dark state analysis}
\label{sec:dark_states}
As thoroughly explained in \cite{Thiel2020}, the presence of a detector in a finite-dimensional quantum system, such as ours, splits the Hilbert space into two parts: a \textit{bright subspace}, whose states are guaranteed to be eventually detected, ($P_{\rm det}(\infty)=1$), and a \textit{dark subspace}, whose states are never detected, not even after an infinite number of detection attempts ($P_{\rm det}(\infty) = 0$). 
A state that lies in the dark subspace is orthogonal (and remains orthogonal under the action of the evolution operator) to the detection site $|\delta \rangle$. 
A state that satisfies these properties---i.e., it is stationary with respect to the evolution and detection attempts---is called a \textit{dark state}.
The total detection probability $P_{\rm det}(\infty)$ is the squared modulus of the projection of the initial state $\vert\psi_0\rangle$ on the \textit{bright space}  \cite{Thiel2020}. Accordingly, whenever $|\psi_0\rangle$ has a nonzero projection on the \textit{dark space}, then $P_{\rm det}(\infty)<1$, meaning that we are not guaranteed to detect the walker even with infinite measurements. Understanding when dark states arise and characterizing them is therefore important to optimize the detection probability.

Considering the time evolution in Eq. \eqref{eq:free_time_evol}, a dark state can be built from the energy eigenstates \eqref{eq:eigvec_eigval},
\begin{equation}
    |\gamma_{n,m}\rangle = \sqrt{\frac{N}{2}} \Big( \langle \delta|\lambda_n\rangle|\lambda_m\rangle -\langle\delta|\lambda_m\rangle|\lambda_n\rangle \Big),
    \label{eq:dark_state_important_example}
\end{equation}
with $m \neq n$. Letting it evolve with $\hat{U}(\tau)$, we get the state
\begin{align}
    \hat{U}(\tau)|\gamma_{n,m}\rangle = \sqrt{\frac{N}{2}} 
    \Big( e^{-i\lambda_m \tau}\langle \delta|\lambda_n\rangle &|\lambda_m\rangle\nonumber\\
    - e^{- i\lambda_n \tau} \langle\delta |\lambda_m\rangle &|\lambda_n\rangle \Big),
\end{align}
which is orthogonal to the detection site $|\delta\rangle$ if 
\begin{equation}
   \lambda_m \tau \equiv \lambda_n \tau \pmod{2\pi} ,
   \label{eq:mod_phase_matching}
\end{equation}
in which case the system never visits $\ket{\delta}$, resulting in $P_{\rm det}(\infty)=0$.
If no pairs of eigenvalues $\lambda_m,\lambda_n$ satisfy the condition \eqref{eq:mod_phase_matching} and $\langle \lambda_j|\delta\rangle \neq  0$ $\forall j$, then the Hilbert space is \textit{fully bright}, and there are no dark states \cite{Thiel2020}. 
Note that when Eq. \eqref{eq:mod_phase_matching} holds true, we have that $[\mathbb{I}-\hat{D}]\hat{U}(\tau)|\gamma_{n,m}\rangle\,=\,e^{-i\lambda_m \tau}|\gamma_{n,m}\rangle$, i.e., 
these dark states are eigenstates of the Perron-Frobenius operator $\hat{O}$ and are the only ones with unit-modulus eigenvalue
\footnote{The time-evolution operator $\hat{U}(\tau)$ in Eq. \eqref{eq:free_time_evol} is unitary, thus preserves the norm and has unit-modulus eigenvalues. Accordingly, for the Perron-Frobenius operator \eqref{eq:PF_operator} to have an eigenvalue of modulus 1 with corresponding eigenstate $|\mu\rangle$, it must be that $\hat{D}\hat{U}(\tau)|\mu\rangle = 0$; otherwise, the non-unitary operator $(\mathbb{I} - \hat{D})$ would reduce the norm of $\hat{U}(\tau)|\mu\rangle$, preventing a unit-modulus eigenvalue of $\hat{O}$. In other words, $|\mu\rangle$ is eigenstate of $\hat{O}$ with unit-modulus eigenvalue only if  $\hat{O}\vert \mu \rangle =\hat{U}(\tau)\vert \mu \rangle$, meaning that $\vert \mu \rangle$ must be an eigenstate of $\hat{U}(\tau)$ orthogonal to $\vert \delta \rangle$.}.
In our model the condition $\langle \lambda_j|\delta\rangle \neq 0 \;\forall j$ is satisfied by any localized state $|\delta\rangle$ [see Eq. \eqref{eq:eigvec_eigval}], as our detection states, while the condition \eqref{eq:mod_phase_matching} can be satisfied in two ways:

(i) When the Hamiltonian's spectrum is degenerate, then Eq. \eqref{eq:mod_phase_matching} is satisfied  $\forall\tau$ for each degenerate level. Degeneracies in the spectrum \eqref{eq:eigvec_eigval} arise only for $\phi = 0$ and $\phi = \pm\pi/N$, and the properties of the resulting dark states depend on the parity of $N$.

For even $N$, when $\phi =0$ there exist $\frac{N}{2} - 1$ two-fold  degenerate levels. In this case, all the stationary dark states constructed from the pairs of eigenstates in each degenerate level are orthogonal to the initial state. Therefore, they neither contribute to the dynamics of the system nor affect the detection probability.
When $\phi = \pm\pi/N$ there exist $N/2$ two-fold degenerate levels. It can be proved (see Appendix~\ref{app:PF_even_N}) that the initial state lies in the dark subspace spanned by the stationary dark states constructed from these degenerate levels. As a consequence, in this case the walker is never detected in $|\delta\rangle$, yielding $P_{\rm det}(\pm\pi/N)=0$ in Fig. \ref{fig:Pdet_density_cuts}(a).

For odd $N$, calculations analogous to those in Appendix~\ref{app:PF_even_N} reveal that
for $\phi = 0, \pm \pi/N$ the total overlap of the initial state with the dark space is $1/2$, thus explaining the minima of the detection probability for these values of $\phi$ in Fig. \ref{fig:Pdet_density_cuts}(b). For such values of $\phi$, there exist
$(N-1)/2$ two-fold degenerate levels, each contributing a dark state that, together, form the basis spanning the dark space.

(ii) Using any nondegenerate couple of eigenvalues $\lambda_m \neq \lambda_n$, there may exist proper combinations of $\tau$ and $\phi$ that lead to $\lambda_m \tau = \lambda_n \tau + 2k\pi$ for some $k\in\mathbb{Z}$.
Plugging the eigenvalues \eqref{eq:eigvec_eigval} in the latter expression, i.e., in Eq. \eqref{eq:mod_phase_matching}, we derive the condition relating $\tau$ and $\phi$ under which dark states appear
\begin{equation}
   \tau_{\rm dark} = \frac{k\pi}{ 2 \sin \left(\frac{\pi  (m-n)}{N}\right) \sin \left(\phi-\frac{\pi 
  (m+n)}{N} \right)},
  \label{eq:tau4darkstates}
\end{equation}
which establishes the lower bound $\tau_{\rm dark} \geq \pi/2$ for the presence of dark states arising from nondegenerate levels.

Such a lower bound is exactly attained, $\tau_{\rm dark}=\pi/2$, for even $N$ when $\phi=0$ (with $k=1$, $m=0$, and $n=N/2$).
In all other cases, the first dark state associated with nondegenerate levels arises at a detection period $\tau_{\rm dark} \to\pi/2$ as $N\to\infty$, manifesting as a dark-state line at $\tau \to\pi/2$ in the parameter space $(\phi,\tau)$. This behavior follows from the fact that, as $N$ increases, the discrete arguments of the sine functions in Eq. \eqref{eq:tau4darkstates} become denser in the interval $[0,2\pi)$, so that pairs $(m,n)$ exist for which the product of the two sine functions is arbitrarily close to 1.
In general, for finite $N$, the solutions to Eq. \eqref{eq:tau4darkstates} manifest as low-detection-probability curves in the $(\phi,\tau)$-space stemming from the presence of dark states which overlap with the initial state, as clearly illustrated in Figs. \ref{fig:Pdet_density_cuts}(a,b), including the nearly straight line at the threshold $\tau^\ast \approx 1.58 \gtrsim \pi/2$.
In the region $\tau\geq\pi/2$, the dark-state curves become denser as $N$ increases, because the number of dark states from nondegenerate levels is of the order $O(N^2)$, and also as $\tau$ increases (see Appendix~\ref{app:density_grows_with_tau}).

In conclusion, the analysis of dark states provides an operational definition of the threshold in the detection period as $\tau^\ast \equiv \min_{m,n,\phi}\tau_{\rm dark}$, where $\tau_{\rm dark}$ is defined in Eq. \eqref{eq:tau4darkstates} and $m,n$ label nondegenerate energy levels. Accordingly, $\tau^\ast \to \pi/2^+$ as $N\to \infty$. In the region $\tau<\tau^\ast$, results qualitatively differ depending on $N$ and $\phi$. For even $N$, dark states at $\phi=0$ do not affect the evolution of $|\psi_0\rangle$ yielding $P_{\rm det}(\infty)=1$, while for $\phi = \pm \pi/N$ the initial state is dark, yielding $P_{\rm det}(\infty)=0$; for odd $N$, the initial state has a finite overlap with dark states at $\phi = 0,\pm \pi/N$, yielding $P_{\rm det}(\infty)=1/2$. For $\phi\neq0,\pm \pi/N$, the Hilbert space is fully bright in this region, and thus $P_{\rm det}(\infty)=1$ for both even and odd $N$ (this is not fully captured in Fig. \ref{fig:Pdet_density_cuts}(a,b) due to the finite observation time $T$).
In the region $\tau > \tau^\ast$, the number of dark states increases as $N$ and $\tau$ increase, yielding denser low-detection-probability curves in the $(\phi,\tau)$-space.

\section{Asymptotic dynamics time scale and finite-time effects}
\label{sec:asymptotic}
While the Perron-Frobenius analysis provides useful insights into the asymptotic dynamics of the system in the long-time limit, experimental scenarios are necessarily constrained to finite observation times $T$ (e.g., we assumed $T=200$ in Fig. \ref{fig:Pdet_density_cuts}) which may not correspond to such regime. This naturally raises the questions: over what timescale 
$t_{\rm as}$ does the asymptotic behavior emerge, and how accessible is it? How predictive is the Perron-Frobenius analysis for the optimal parameters at finite time?

\begin{figure}[!t]
    \centering
    \includegraphics[width=0.9\columnwidth]{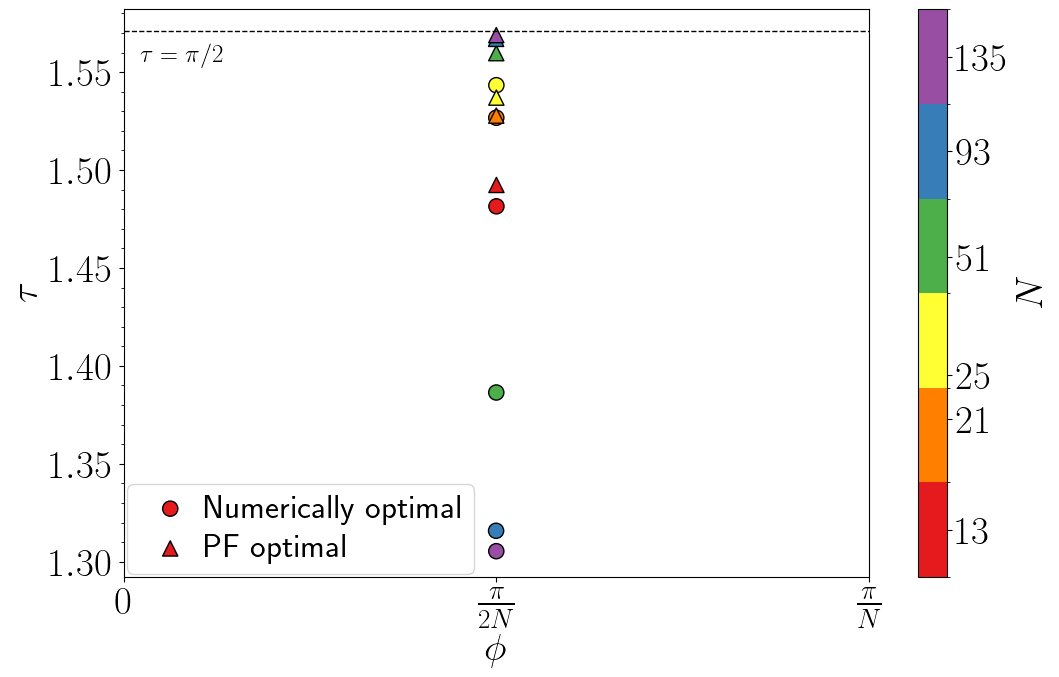}
    \caption{Comparison between the Perron-Frobenius (PF) optimal parameters $(\phi_{\rm PF},\tau_{\rm PF})$, obtained by minimizing $\vert \mu_{\rm PF} \vert$, and the numerically-optimal parameters $(\phi_{\rm opt},\tau_{\rm opt})$, obtained by maximizing the detection probability, as functions of odd $N$ at finite observation time $T=200$. While the optimal phase is $\phi_{\rm opt}=\phi_{\rm PF}$ regardless of $N$, the optimal detection periods deviate from each other as $N$ increases: $\tau_{\rm PF} \to \pi/2$ whereas $\tau_{\rm opt}$ decreases.}
    \label{fig:PF_vs_num_optimal}
\end{figure}

\begin{figure*}[!t]
    \centering
    \includegraphics[width=0.7\textwidth]{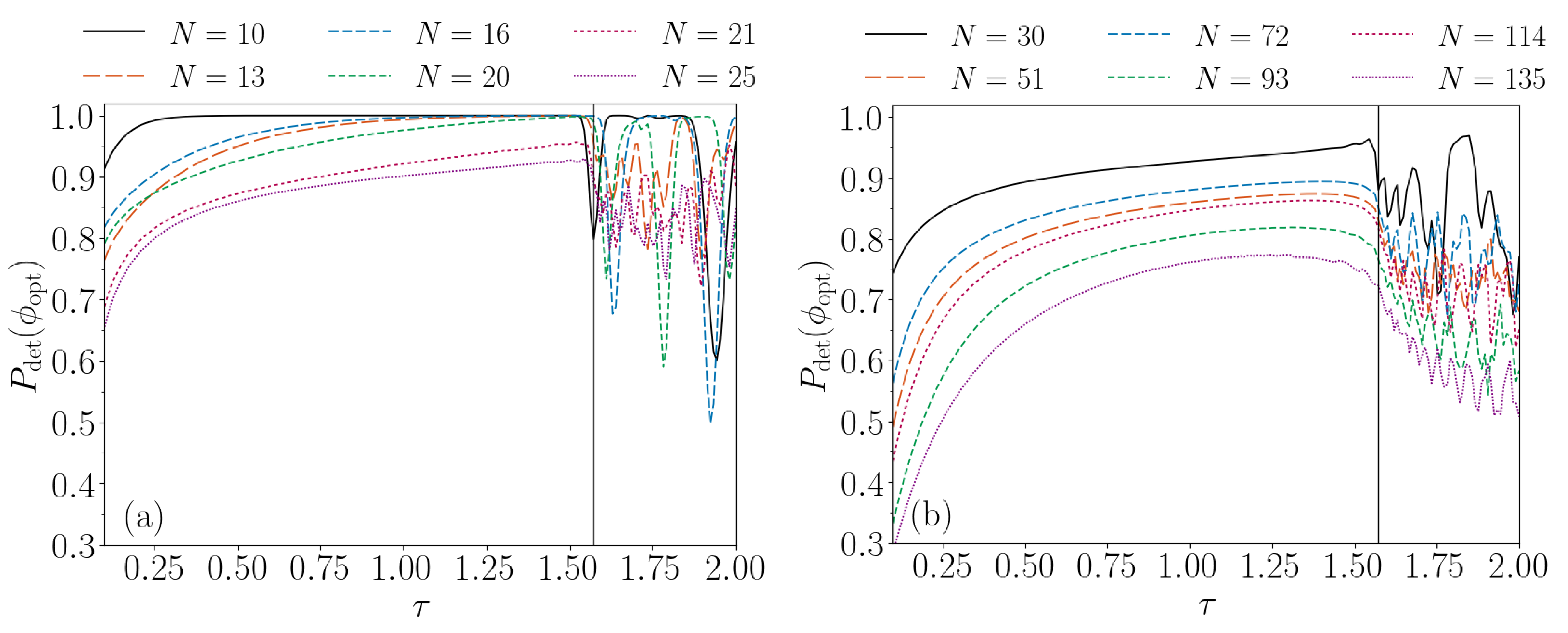}
    \caption{Detection probability as a function of $\tau$ evaluated at fixed $\phi=\phi_{\rm opt}$ and $T=200$ for various $N$.
    (a) Small $N$ for which the system reaches its asymptotic behavior within the observation time $T$. (b) High $N$ for which $T\ll t_{\mathrm{as}}$.
    The reference value $\tau=\pi/2$ is shown as vertical line.
    The parameters used are $\delta=N/2$ and $\phi_{\rm opt}=0$ for even $N$, $\delta=(N-1)/2$ and $\phi_{\rm opt}=\pi/2N$ for odd $N$.
    }
    \label{fig:Pdet_tagli}
\end{figure*}

At finite observation times $T$, the extent to which the Perron-Frobenius optimal parameters---$\phi_{\rm PF}$ and $\tau_{\rm PF}$ which minimize $|\mu_{\rm PF}|<1$---are consistent with the numerically optimal ones---$\phi_{\rm opt}$ and $\tau_{\rm opt}$ which maximize $P_{\rm det}$---depends on how close $T$ is to the asymptotic time scale $t_{\rm as}$.
If $T \gg t_{\rm as}$, then the asymptotic regime holds true and $\phi_{\rm opt}=\phi_{\rm PF}$ and $\tau_{\rm opt}=\tau_{\rm PF}$ maximize the detection probability by minimizing the survival probability \eqref{eq:surv_prob}, being $\vert \mu_j \vert \leq \vert \mu_{\rm PF} \vert < 1\,\forall j$ by definition.
If $T \lesssim t_{\rm as}$, then finite-time effects arise and affect the accuracy of the predictions. To illustrate this, it is instructive to examine how the optimal parameters vary as a function of the system size $N$ at fixed observation time $T$ (see Fig. \ref{fig:PF_vs_num_optimal}). At low values of $N$, $\phi_{\rm PF} = \phi_{\rm opt}$ and $\tau_{\rm PF}\approx \tau_{\rm opt}$, where minor discrepancies in the detection period arise either from the numerical maximization of a flat curve (e.g, for $N=13$) or the emergence of finite-time effects (e.g. for $N=25$) [see also Fig. \ref{fig:Pdet_tagli}(a)].
As $N$ increases, the prediction remains accurate for the optimal phase $\phi_{\rm PF} = \phi_{\rm opt}$ (a Hamiltonian parameter) but becomes progressively less accurate for the detection period, with $\tau_{\rm PF}\to \pi/2^-$ and $\tau_{\rm opt}$ decreasing.
Consistently with this picture, Fig. \ref{fig:Pdet_tagli} shows that for moderate system sizes $N$ the detection probability  may not saturate to the expected value $P_{\rm det}(\infty)=1$ within the available observation time $T=200$, and Perron-Frobenius optimal detection periods---which approach $\tau = \pi/2$ as $N$ increases---may yield suboptimal results at finite times (see the optimal values $\tau_{\rm opt}$ in Fig. \ref{fig:PF_vs_num_optimal}). The detection probability correctly saturates to the asymptotic value for low system sizes, $N \leq 20$. As is intuitive, the larger the system size, the longer the time it takes the excitation to reach the target, and so the longer the time scale over which the asymptotic behavior emerges and the detection probability saturates.

\begin{figure}[!b]
    \centering
    \includegraphics[width=\columnwidth]{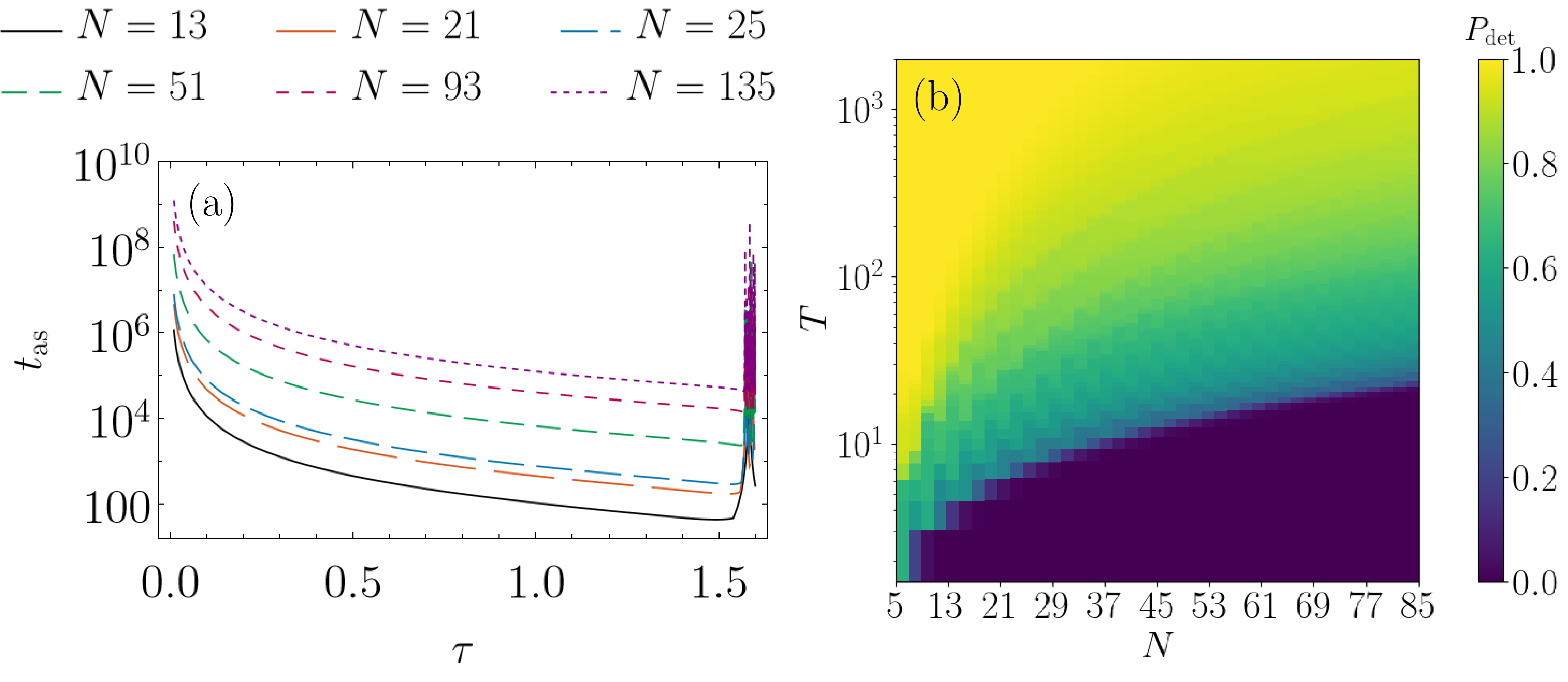}
    \caption{(a) Asymptotic time $t_{\rm as}$ as a function of the detection period $\tau$ for various odd $N$. (b) Density plot of $P_{\rm det}$ as a function of odd $N$ and total observation time $T$, evaluated at $\tau_{\rm PF}$    
    corresponding to each $N$ and at $\delta=(N-1)/2$. In both panels the phase is fixed at $\phi_{\rm opt}=\pi/2N$.}
    \label{fig:asymptotic_tscale}
\end{figure}

The asymptotic time scale is governed by the spectral properties of the Perron-Frobenius operator \eqref{eq:PF_operator}, specifically by the spectral gap $\Delta \equiv 1-|\mu_{\rm PF}|$.
Therefore, the  asymptotic time scale can be estimated as $t_{\rm as} \sim\Delta^{-1}$ \footnote{Analogous results and conclusions can be drawn from considering the inverse of the decay rate of the survival probability, $\min_j\{-2 \ln \vert \mu_j\vert \}$ \cite{Thiel2020}.}.
Considering that $t_{\rm as}$ depends on $N$, $\tau$, and $\phi$, in Fig.~\ref{fig:asymptotic_tscale}(a) we restrict our analysis to $t_{\rm as}$ as function of $\tau$ for different $N$ by fixing $\phi$ at its known optimal value, since results indicate that $\phi_{\rm opt} = \phi_{\rm PF}$ regardless of the finite observation time (see Fig. \ref{fig:PF_vs_num_optimal}).
In the small-$\tau$ limit, $t_{\rm as}$ diverges due to the quantum Zeno effect, which prevents the walker from reaching the measurement site. As $\tau$ increases, $t_{\rm as}$ decreases monotonically, reaching a minimum at $\tau_{\rm PF}$ (where $|\mu_{\rm PF}|$ is minimized), before rising sharply. Beyond the threshold $\tau^*$, $t_{\rm as}$ reflects the system's sensitiveness to small changes in $\tau$ and $\phi$ [see also Fig. \ref{fig:Pdet_density_cuts}(b)]. This trend persists with increasing $N$, which leads to larger $t_{\rm as}$.
Clearly, the position of the minimum of $t_{\rm as}$ approaches $\pi/2$ as $N$ increases, since the minimum of $|\mu_{\rm PF}|$ occurs at $\tau_{\rm PF}\to \pi/2$ as $N\to \infty$ (see Fig. \ref{fig:PF_vs_num_optimal}). Minimizing $\vert\mu_{\rm PF}\vert$ is therefore crucial, as it reduces the time scale over which the asymptotic behavior emerges, making it observable within finite times.

The detection probability saturates to its asymptotic value over a time scale $t_{\rm as}$ that increases with the system size $N$. In practice, the experimentalist may only require the detection probability to exceed a threshold. Motivated by this, in Fig.~\ref{fig:asymptotic_tscale}(b) we study $P_{\rm det}$ as function of $N$ and total observation time $T$ at $\tau_{\rm PF}$ and $\phi_{\rm PF}=\phi_{\rm opt}$. Although sub-optimal at finite times, we adopt these parameters to complement the analysis initiated in Fig.~\ref{fig:PF_vs_num_optimal}.
As $N$ increases, the observation time $T$ required for $P_{\rm det}$ to become non-negligible and subsequently saturate also increases.
In this regard, $T = 200$ considered in the present work effectively captures the asymptotic dynamics, $T \gtrsim t_{\rm as}$, for $N \leq 20$, while at $N = 21$ $T \sim t_{\rm as}$ yields $P_{\rm det}\lesssim 1$ (see Fig. \ref{fig:Pdet_tagli}). 
Remarkably, achieving $P_{\rm det} \approx 80\%$ requires an observation time $T$ nearly an order of magnitude shorter than that needed for saturation with the Perron-Frobenius (sub-)optimal parameters.

To summarize, the Perron-Frobenius analysis yields optimal parameters for maximizing detection probability in the asymptotic regime ($T \gg t_{\rm as}$), and it remains a valuable tool even in the finite-time regime ($T \lesssim t_{\rm as}$). Finite-time effects stem from the interplay between finite observation time $T$ and system size $N$, causing deviations from the asymptotic behavior. Specifically, for $T \lesssim t_{\rm as}$, the chiral phase is correctly predicted, $\phi_{\rm PF}=\phi_{\rm opt}$, while $\tau_{\rm opt}$ deviates from $\tau_{\rm PF}$, decreasing with $N$ whereas $\tau_{\rm PF} \to \pi/2^-$. Therefore, whenever the finite-time effects are moderate, $\tau_{\rm PF}$ can still serve as initial guess for refining the optimization. Careful consideration is required when these finite-time effects become particularly pronounced. This framework is valid only if the observation time $T$ is sufficiently long relative to the system size $N$, so that excitation transport to the target site can occur. This constrains the regime of finite-time dynamics of interest: recalling that, for a continuous-time quantum walk on the line, the standard deviation of the position is linear with time (ballistic spreading), we require at least $T \propto N$ to have transport
and, in addition, we observe that $t_{\rm as}(\tau_{\rm PF}) \sim N^3$ (see Appendix~\ref{app:t_as_vs_N}).

\begin{figure*}[!t]
    \centering
    \includegraphics[width=0.65\textwidth]{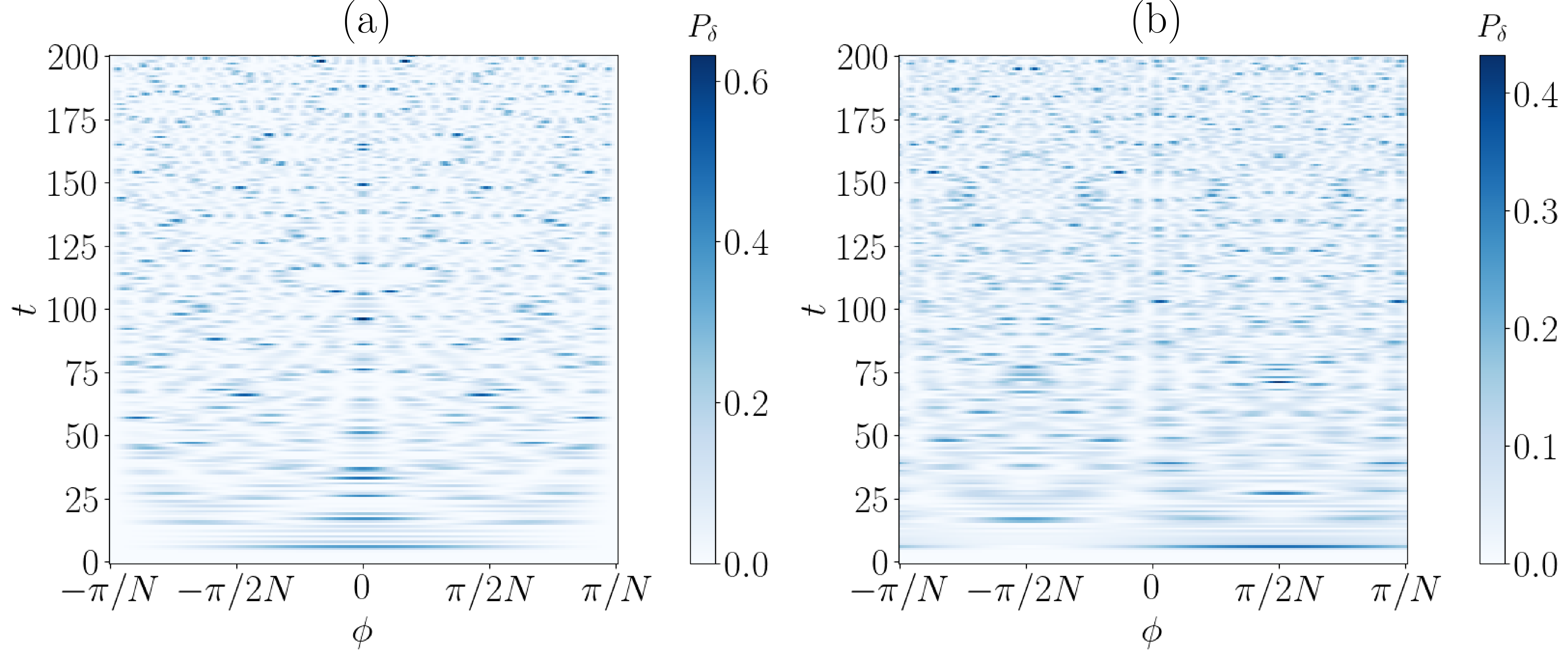}
    \caption{Instantaneous transfer probability $P_\delta$ from site $0$ to $\delta$ as a function of the phase $\phi$ and time $t$ (a) for $N=20$ and $\delta=N/2$ and (b) for $N=21$ and $\delta=(N-1)/2$. 
    }
    \label{fig:Pinst_density}
\end{figure*}

\section{Conclusion}
\label{sec:conclusions}
In this work, we have investigated how chirality and local monitoring jointly enhance excitation transfer, modeled as a continuous-time quantum walk on a ring. We show that optimizing both the chiral phase $\phi$ and the detection period $\tau$ overcomes limitations of purely unitary dynamics, such as destructive interference and dark states. The resulting transfer protocol is both optimal, yielding significantly higher detection probabilities, and robust, requiring no fine-tuning of parameters or tailored initial-state preparations.
Our approach combines two key insights: (i) The identification of dynamically relevant dark states,
and (ii) the spectral analysis of the non-unitary Perron-Frobenius operator to determine optimal parameters. While this analysis is exact in the asymptotic regime, it remains a valuable tool at finite times: it yields the correct optimal phase and provides a useful estimate of the optimal detection time, whose accuracy worsens as the system departs further from the asymptotic regime. 
This approach, which remains effective provided the observation time $T$ grows with the system size $N$,
offers a general framework for enhancing transport in monitored quantum systems beyond the simple model investigated in the present work.

Beyond advancing the understanding of first-detected-passage problems in closed loops, our results offer practical guidelines that can be readily extended beyond quantum transport to the design of efficient quantum protocols in domains where destructive interference and dark states usually limit performance, such as search algorithms \cite{childs2002pra,childs2004pra}, state transfer \cite{kay2011}, quantum routers \cite{bottarelli2023}, and potentially in quantum-walk-based computing \cite{qiang2024}.

Finally, hitting times play a central role in the performance of quantum protocols, as they directly affect how rapidly a target state can be reached or detected. The recent implementation of local monitoring via mid-circuit measurements---either strong \cite{tornow2023prr,wang2024entropy} or weak \cite{heine2025}---demonstrates the growing experimental feasibility of accessing quantum hitting-time statistics on current quantum hardware. In parallel, restart strategies have emerged as a new, powerful control mechanism to further speed up hitting times in monitored quantum systems \cite{yin2023,roy2025pre,yin2025,yin2025pnas}. Together, these advances underscore the relevance of our approach for designing efficient and experimentally viable quantum protocols.

\begin{acknowledgments}
S.F., G.B., and L.R. acknowledge support from INFN through the project `QUANTUM'. S.F. and G.B. acknowledge support from MUR and EU through Grant No. PNRR D.M. 118/2023. This work was done under the auspices of GNFM-INdAM. M.G.A.P. acknowledges partial support from MUR and EU through the projects PRIN22-PNRR QWEST (CUP G53D23006270001), NQSTI-Spoke1-BaC QBETTER (CUP G43C22005120007), NQSTI-Spoke2-BaC QMORE (CUP J13C22000680006).
\end{acknowledgments}

\appendix
\section{Purely-coherent transport under unitary evolution}
\label{app:unitary}
The purely-coherent transport under unitary dynamics is not efficient, as the instantaneous probability $P_\delta(\phi,t)=|\langle \delta \vert \hat{U}(\phi,t) \vert \psi_0\rangle|^2$ typically remains low throughout the evolution, punctuated by narrow, sharp peaks.
This is evident from the comparison between 
Fig.~\ref{fig:Pinst_density} and Fig.~\ref{fig:Pdet_density_cuts}: Fig.~\ref{fig:Pinst_density}
shows $P_\delta$ as a function of the phase $\phi$ and time $t$, revealing that its values are consistently and significantly lower over the entire parameter range considered than those of the detection probability $P_{\rm det}$ in Fig.~\ref{fig:Pdet_density_cuts}.

\section{Perron-Frobenius analysis for even $N$}
\label{app:PF_even_N}
The Perron-Frobenius analysis 
in Sec. \ref{sec:PF_analysis}
focused on the case of odd $N$. Here we discuss the case of even $N$, which needs to be considered more carefully. The main differences between even and odd $N$ root back to the degeneracies of the energy spectrum in 
Eq. \eqref{eq:eigvec_eigval}.
For odd $N$, all the energy levels but one---the highest ($j=0$)---have a two-fold degeneracy when $\phi = 0$. These degeneracies are lifted when $\phi\neq 0$. For even $N$, instead, all the energy levels but two---the lowest ($j=N/2$) and the highest ($j=0$)---have a two-fold degeneracy when $\phi = 0$. Nonzero phases lift these degeneracies except when $\phi = \pm \pi/N$, in which case all the energy levels, including the lowest and the highest, have a two-fold degeneracy.  Following \cite{Thiel2020}, we can construct a dark state from each degenerate level $\lambda_j$ as follows
\begin{equation}
    |\gamma_{j}\rangle = \sqrt{\frac{N}{2}} \Big( \langle \delta|\lambda_{j,2}\rangle|\lambda_{j,1}\rangle -\langle\delta|\lambda_{j,1}\rangle|\lambda_{j,2}\rangle \Big).
\end{equation}

\begin{figure}[!t]
    \centering
    \includegraphics[width=\columnwidth]{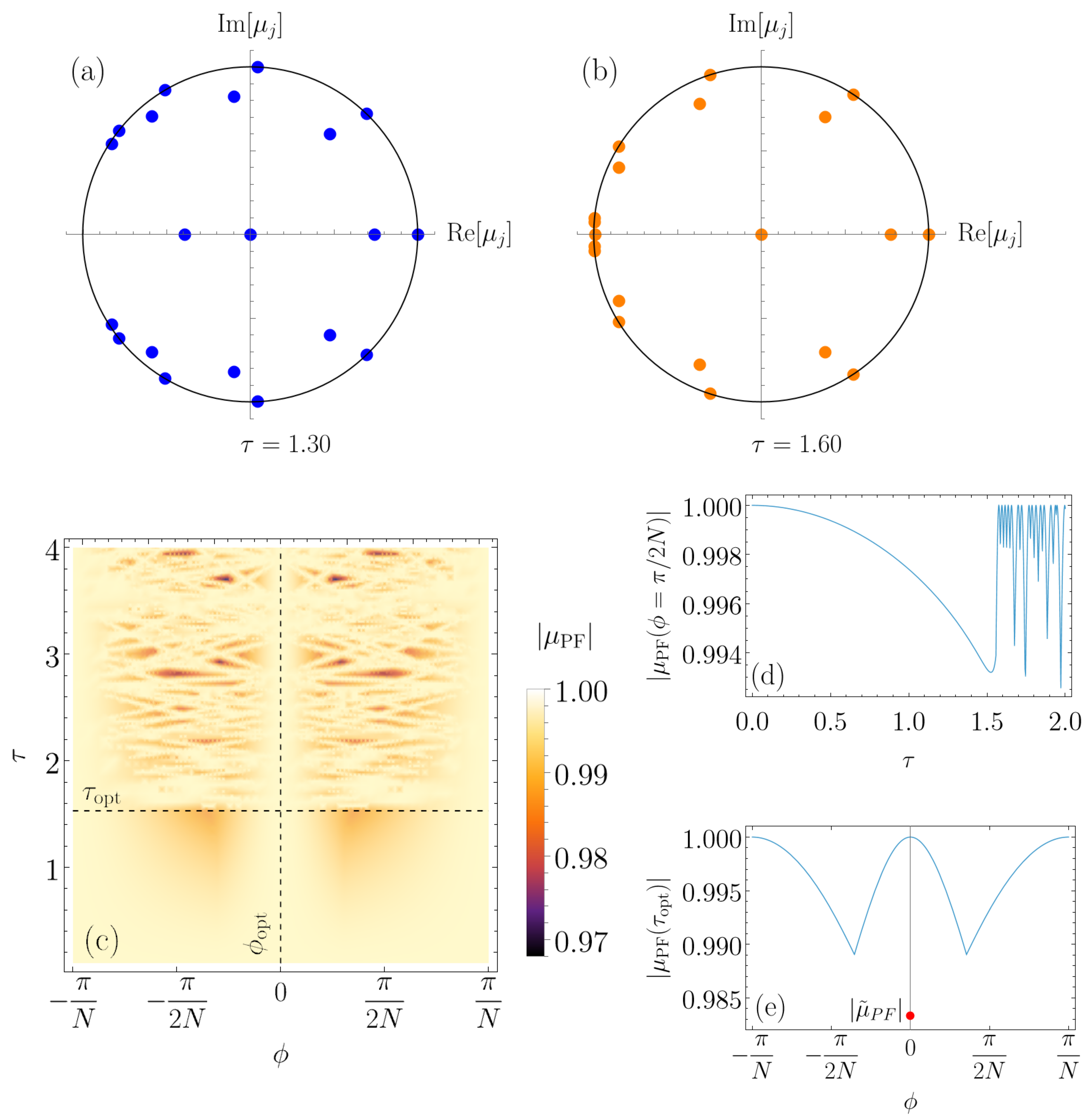}
    \caption{Eigenvalues of the Perron-Frobenius operator \eqref{eq:PF_operator} on the unit circle for two representative values of the detection period, (a) $\tau<\tau^*$ and (b) $\tau>\tau^*$ at $\phi = 0$.   
    (c) Density plot of the largest-modulus PF eigenvalue $|\mu_{\rm PF}|$ as a function of $\phi$ and $\tau$. (d) Curve $|\mu_{\rm PF}|$ as a function of $\tau$ at fixed $\phi=\pi/2N$. (e) Curve $|\mu_{\rm PF}|$ as a function of $\phi$ at fixed $\tau=1.53$. The red circle denotes the subleading PF eigenvalue $\vert \tilde{\mu}_{\rm PF} \vert$, which is the relevant one for the asymptotic dynamics at $\phi = 0$ (the initial state has zero overlap with dark states).  Other parameters are: $N=20$, $\delta=10$. 
    }
    \label{fig:maxmodPF_Neven}
\end{figure}

For $\phi = 0$, the dark states read \cite{Thiel2020}
\begin{equation}
    |\gamma_{j}\rangle = \sqrt{\frac{2}{N}} \sum_{k=0}^{N-1} \sin\left(\frac{2 \pi jk }{N}\right)\vert k \rangle
\end{equation}
with $j = 1,\ldots,N/2-1$, and have no overlap with the initial state, $\langle 0 \vert \gamma_j \rangle = 0 \, \forall j$. As a result, these states do not contribute to the walker's dynamics and can be considered irrelevant in the evolution. These dark states are eigenstates of the Perron-Frobenius operator with unit-modulus eigenvalue
[see Fig.~\ref{fig:maxmodPF_Neven}(a,b)]. 
Consequently, the asymptotic dynamics relevant to excitation transfer will be governed by the subleading eigenvalue, defined as $\tilde{\mu}_{\rm PF} \equiv \max_{\vert \mu_k \vert <1} (\mu_k)$, i.e., the largest eigenvalue in modulus strictly less than one.

For $\phi = - \pi/N$, instead, the dark states read
\begin{equation}
    |\gamma_{j}\rangle = \frac{1}{\sqrt{2N}} \sum_{k=0}^{N-1} \left(e^{i\frac{2 \pi jk }{N}}+e^{-i\frac{2 \pi (j+1)k }{N}}\right)\vert k \rangle 
\end{equation}
with $j=0, \ldots, N/2-1$, and always overlap with the initial state, $\langle 0 \vert \gamma_j \rangle = \sqrt{2/N} \, \forall j$. The initial state turns out to be completely dark for $\phi = - \pi/N$, as it is the equal superposition of all the $N/2$ dark states forming a basis for the dark subspace,
$|0\rangle = \sqrt{2/N} \sum_{j=0}^{\frac{N}{2}-1} |\gamma_{j}\rangle$.
The case $\phi = \pi/N$ can be treated analogously. Therefore, $P_{\rm det} (n\to \infty)=1$ if $\phi = 0$, and $P_{\rm det} (n\to \infty)=0$ if $\phi = \pm \pi/N$ \cite{zimboras2013scirep}.

The study of the largest-modulus Perron-Frobenius eigenvalue 
$\mu_{\rm PF}$ is shown in Fig.~\ref{fig:maxmodPF_Neven}(c).
Similarly to the case  of odd $N$, $|\mu_{\rm PF}|$ smoothly attains a local minimum (which we identify as $\tau_{\rm opt}$ in the asymptotic regime) in the interval $\tau<\tau^\star$, and then rapidly oscillates [Fig. \ref{fig:maxmodPF_Neven}(d)].
As for the dependence on the phase $\phi$,
$\mu_{\rm PF}$ shows two symmetric minima at $\phi\ne 0$,
in contrast with the numerically optimal phase $\phi_{\rm opt}=0$ that we expect for even $N$ (see Fig. \ref{fig:Pdet_density_cuts}).
On the other hand, as discussed above, dark states built using degenerate energy levels are irrelevant for excitation transfer at $\phi=0$. At $\phi = 0$, one should thus consider the modulus of the subleading eigenvalue, $|\tilde{\mu}_{\rm PF}|$ [red dot in 
Fig. \ref{fig:maxmodPF_Neven}(e)], which is lower than the two local minima of $|\mu_{\rm PF}|$ when $\phi\ne 0$. 
This is reflected in the fact that the detection probability attains a higher maximum at $\phi=0$ than at the local maxima occurring at $0<|\phi|<\pi/N$.
We therefore conclude that the optimal phase for even $N$ is $\phi = 0$.

\section{Density of dark states increases with $\tau$}
\label{app:density_grows_with_tau}
We prove that the density of dark states---i.e., the number of dark states in a certain interval $[\phi, \phi +\Delta\phi] $ at fixed $\tau$---grows with $\tau$ in the region $\tau>\pi/2$.
As discussed in Sec. \ref{sec:dark_states},
outside of the degenerate case, dark states correspond to solutions of 
Eq. \eqref{eq:tau4darkstates}.
Graphically, these solutions are the intersections of two functions, $y_1 = k\pi$ and 
$y_2 = 2 \tau \sin \left[\pi (m-n)/N\right]\sin \left[\phi-\pi(m+n) /N\right]$. On the plane $(\phi,y)$, the function $y_1$ represents horizontal lines at integer multiples of $\pi$; at fixed $\tau,m,n,N$, the function $y_2$ is a sinusoidal function in $\phi$, with $\tau$ acting as an amplification factor, provided that $m,n$ satisfy $\sin \left[\pi (m-n)/N\right]\neq0$.
It is straightforward to see that the number of horizontal lines $y_1=k\pi$ crossed by the function $y_2$ grows in any interval $[\phi, \phi +\Delta\phi] $ as $\tau$ increases, as was to be demonstrated. 
In other words, for larger values of $\tau$, a finer tuning on $\phi$ is needed to avoid dark states, solutions of $y_1=y_2$. Searching for a robust transfer protocol, this further motivates our choice of restricting to $\tau<\tau^*$.

\begin{figure}[!t]
    \centering
    \includegraphics[width=0.9\columnwidth]{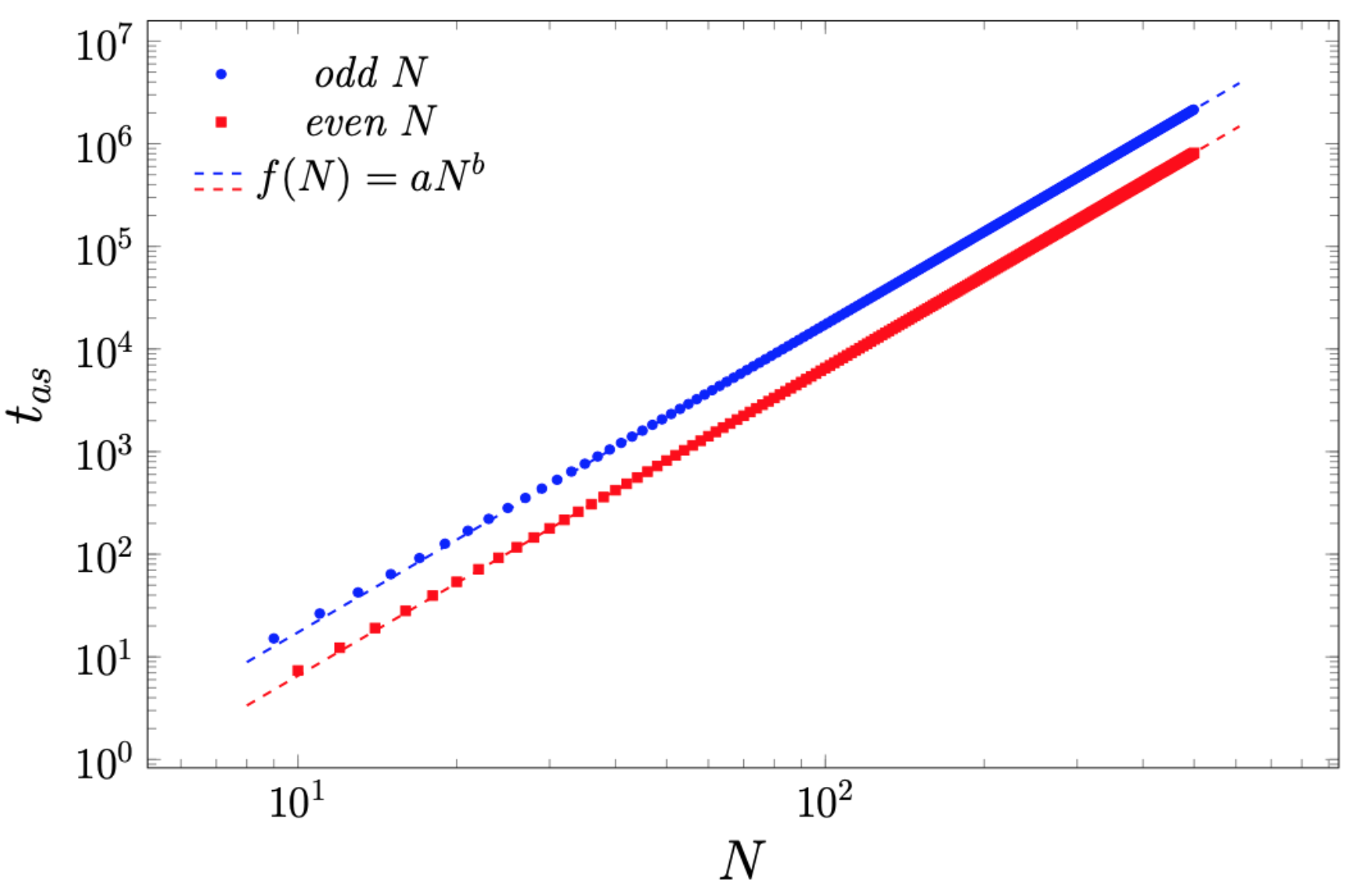}
    \caption{Asymptotic time scale $t_{\rm as}$ as a function of the system size $N$ for odd (blue circles) and even (red squares) values. A power-law fit $f(N) = a N^b$ shows that the scaling is nearly cubic, $t_{\rm as} \sim N^3$, independent of the parity of $N$.     
    In particular, it returns exponents $b^{\rm (odd\,N)} = 2.999434 \pm 0.000017$ and $b^{\rm (even\,N)} = 2.9988 \pm 0.0003$, close to 3. Differences due to the parity of $N$ are reflected in the proportionality constants, $a^{\rm (odd\,N)} = 0.0173855 \pm 0.0000018$ and $a^{\rm (even\,N)} = 0.006562 \pm 0.000013$, which differ by a factor $\approx 2.65$.    
    }
    \label{fig:scaling_Tas_vs_N}
\end{figure}

\section{Relation between asymptotic time scale and system size}
\label{app:t_as_vs_N}
Finite observation times $T$ may not fall within the asymptotic time scale $t_{\rm as}$, where the Perron-Frobenius analysis holds and is well suited to investigate the asymptotic dynamics of the system. The asymptotic time scale clearly depends on the system size $N$: larger systems require longer times for the excitation to reach the target state and for the dynamics to enter the asymptotic regime. In Fig. \ref{fig:scaling_Tas_vs_N} we show that the asymptotic time scale, defined as the inverse of the spectral gap of the Perron-Frobenius operator $t_{\rm as} \equiv (1-|\mu_{\rm PF}|)^{-1}$, essentially scales as $t_{\rm as} \sim N^3$, independently of the parity of $N$.
Analogous scaling is observed when the asymptotic time scale is defined as the inverse of the decay rate of the survival probability, $t_{\rm as} \equiv (\min_j\{-2 \ln \vert \mu_j\vert \})^{-1} \sim N^3$ \cite{Thiel2020}. The knowledge of this scaling allows one to estimate the lower bound on the observation time necessary for optimizing excitation transfer, as predicted by Perron-Frobenius analysis in the asymptotic regime.

\bibliography{biblio}

\end{document}